\documentclass[preprint,proceedings]{rmaa}

\SetYear{2002}
\SetConfTitle{Galaxy Evolution: Theory and Observations}

\title{TESIS - The TNG EROs Spectroscopic Identification Survey}

\author{
  P. Saracco\altaffilmark{1}, M. Longhetti\altaffilmark{1}, 
P. Severgnini\altaffilmark{1}, R. Della Ceca\altaffilmark{1}, 
 R. Bender\altaffilmark{2}, N. Drory\altaffilmark{2}, 
G. Feulner\altaffilmark{2}, F. Ghinassi\altaffilmark{3},
U. Hopp\altaffilmark{2}, F. Mannucci\altaffilmark{4}, 
C. Maraston\altaffilmark{2}}

\altaffiltext{1}{Osservatorio Astronomico di Brera, Via E. Bianchi 46,
23807 Merate (LC), Italy (saracco@merate.mi.astro.it).}
\altaffiltext{2}{Universitats-Sternwart Munchen, Munchen, Germany.}
\altaffiltext{4}{IRA-CNR, Firenze, Italy.}
\altaffiltext{3}{Centro Galileo Galilei, La Palma, Spain}
\suppressfulladdresses

\listofauthors{P. Saracco, M. Longhetti, P. Severgnini, R. Della Ceca, 
R. Bender, N. Drory, G. Feulner, F. Ghinassi, U. Hopp, F. Mannucci, 
C. Maraston}
\indexauthor{Saracco, P.}
\indexauthor{Longhetti, M }
\indexauthor{Severgnini, P. }
\indexauthor{Della Ceca, R. }
\indexauthor{Mannucci, F. }
\indexauthor{Ghinassi, F.}
\indexauthor{Drory, N. }
\indexauthor{Feulner, G.}
\indexauthor{Bender, R.}
\indexauthor{Maraston, C.}
\indexauthor{Hopp, U.}

\addkeyword{Galaxies: evolution}
\addkeyword{Galaxies: elliptical}

\begin{document}
\maketitle 

\boldabstract{The epoch at which massive  
galaxies (M$_{star}>10^{11}$M$_\odot$) have assembled provides crucial 
constraints on the current galaxy formation and evolution models.
The LCDM hierarchical merging model predicts that
massive galaxies are assembled through mergers 
of pre-existing disk galaxies at z$\le1.5$
 (Kauffmann \& Charlot 1998; Cole et al. 2000).
In the alternative view  massive ellipticals formed 
at z$>3$ in a  single episode of star formation and
follow  a pure luminosity evolution (PLE).
}
The number
density of M$_{star}>10^{11}$M$_\odot$ galaxies predicted in the LCDM model
by Baugh et al. (2002) is $\sim10^{-5}h^3$Mpc$^{-3}$ at $z\simeq1$.
which gives an upper limit to the surface density of massive galaxies
of 0.02 gal/arcmin$^2$ in the redshift range $1.2<z<1.5$.
 Given this low predicted
probability, searching for massive galaxies at $z\ge1.2$ and estimating
their number density provide strong observational constraints on models.
With the aim at identifying massive evolved  galaxies at 
$z\ge1.2$,  we started a near-IR very low resolution 
spectroscopic survey of a complete sample of K'$<18.5$ EROs (R-K'$>5$).
Indeed, K$\sim18.5$ is the expected apparent K-band magnitude of a  
M$_{star}=10^{11}$M$_\odot$ galaxy at $z\sim1$. 
 The sample (15 EROs at K'$<18$ and 52 at K'$<18.5$) has been 
selected from 
two no-adjacent wide areas of about 180 arcmin$^2$ each of the Munich 
Near-IR Cluster Survey (MUNICS, Drory et al. 2001).
The observations are based on the Amici prism dispersing element
  mounted at the near-IR camera NICS of the Telescopio Nazionale 
Galileo (TNG). 
This observing mode provides the spectrum from 8500-24000 \AA~
in one shot  with a nearly constant resolution 
($\lambda/\Delta\lambda\sim50$) and a high throughput ($\sim80\%$). 
For these reasons the Amici data result very efficient in 
estimating redshift of old stellar systems in
 the $1.2<z<2$ range through the detection of the Balmer break.  
On the other hand, both the low  resolution  and the low S/N ($\sim3$)
achievable at these magnitudes make the detection and the estimate of 
absorption/emission lines uncertain.
During the observing run of March 2002, we carried out near-IR 
spectra of 4 EROs with K$<18$ under unclear sky  and  
 poor seeing conditions (1.5-2.5 arcsec).
We clearly detect in two of them the sharp decrease of the continuum 
expected at 4000 \AA~. The remaing two spectra have a lower S/N  
and the analysis is still in progress. 
The two analyzed spectra are shown in Figure 1. The decrease happens
in the range 0.88-1.0 $\mu$m in both of them constraining 
the ERO S7F5-45 at $1.3<z<1.35$ and the ERO  S7F5-254 at $1.25<z<1.3$.
Given the redshift and the absolute K-band magnitudes derived
(M$_K\simeq-27$) we estimate a lower limit to their
stellar mass of M$_{star}\ge10^{11}$M$_\odot$ under the very conservative
assumption of M$_{star}$/L$_K$=0.1 [M/L]$_\odot$.

\begin{figure}[!t]
  \includegraphics[width=\columnwidth,height=8cm]{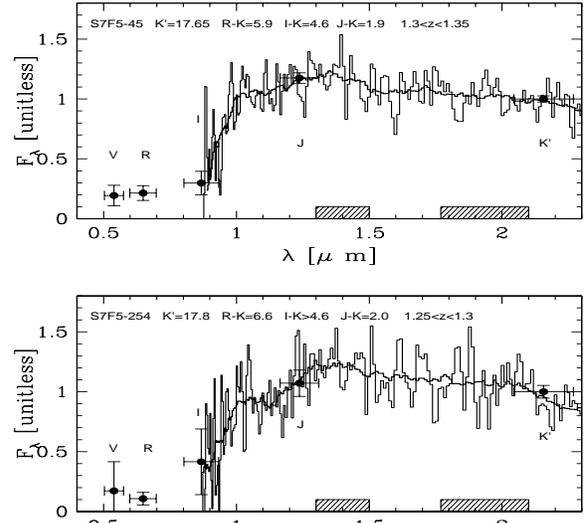}
\vskip -1truecm
  \caption{NICS-Amici low resolution spectra of two 
observed EROs. The  thick histogram is the observed spectrum
heavily smoothed to match the continuum. The spectra are scaled to
the K'-band flux so that F$_K$=1.
The shaded areas represent the atmospheric 
windows characterized by an opacity larger than 80\%.
 The filled symbols are the photometric data in the V, R, I, J and 
K' bands from the MUNICS survey. 
}
  \label{fig:simple}
\end{figure}


\end{document}